\documentstyle[aps,floats,epsf]{revtex}

\def\nbslash{\rlap{\hspace{0.02cm}/}{\bar n}}

\begin{document}
\draft
\twocolumn[\hsize\textwidth\columnwidth\hsize\csname@twocolumnfalse%
\endcsname


\title{\boldmath
Enhanced Non-local Power Corrections to the $\bar B\to X_s\gamma$ Decay Rate
\unboldmath}

\author{Seung J. Lee$^a$, Matthias Neubert$^{a,b}$ and Gil Paz$^c$}

\address{$^a$\,Institute for High-Energy Phenomenology, Laboratory for 
Elementary-Particle Physics,\\ 
Cornell University, Ithaca, NY 14853, U.S.A.\\
$^b$\,Institut f\"ur Physik (ThEP), Johannes Gutenberg-Universit\"at, 
D--55099 Mainz, Germany\\
$^c$\,School of Natural Sciences, Institute for Advanced Study, 
Princeton, NJ 08540, U.S.A.}
\maketitle

\begin{abstract}
A new class of enhanced non-perturbative corrections to the inclusive
$\bar B\to X_s\gamma$ decay rate is identified, which contribute first
at order $\Lambda/m_b$ in the heavy-quark expansion and cannot be
described using a local operator product expansion. Instead, these
effects are described in terms of hadronic matrix elements of
non-local operators with component fields separated by light-like
distances. They contribute to the high-energy part of the
photon-energy spectrum but do not reduce to local operators when an
integral over energy is taken to obtain the total inclusive decay
rate. The dominant corrections depend on the flavor of the $B$-meson
spectator quark and are described by tri-local four-quark
operators. Their contribution is estimated using the vacuum insertion
approximation. The corresponding uncertainty in the total decay rate
is found to be at the few percent level.  This new effect accounts for
the leading contribution to the rate difference between $B^-$ and
$\bar B^0$ mesons.
\end{abstract}

\pacs{Preprint: CLNS~06/1978, MZ-TH/06-17}]
\narrowtext

\section{Introduction}

Precision studies of inclusive $B$-meson decays are a cornerstone of
quark flavor physics. Detailed measurements of various kinematical
distributions in the semileptonic decays $\bar B\to X\,l\,\bar\nu$,
when combined with elaborate theoretical calculations, provide the
currently most precise measurements of the elements $|V_{cb}|$ and
$|V_{ub}|$ of the quark mixing matrix (see \cite{Lange:2005yw} for a
comprehensive recent analysis of charmless inclusive decays). Studies
of the rare decays $\bar B\to X_s\gamma$ and $\bar B\to X_s\,l^+ l^-$
allow sensitive tests of the flavor sector and provide constraints on
extensions of the Standard Model.

The theoretical description of inclusive $B$-meson decay rates is
based on the operator product expansion (OPE)
\cite{Blok:1993va,Manohar:1993qn}, by which total decay rates can be
expressed in terms of forward $B$-meson matrix elements of local
operators. Only two non-trivial matrix elements appear up to order
$(\Lambda/m_b)^2$ in the expansion, one of which can be extracted from
spectroscopy. The OPE breaks down when one tries to calculate
differential inclusive decay distributions near phase-space
boundaries. A twist expansion involving forward matrix elements of
non-local light-cone operators (so-called shape functions) is then
required to properly account for non-perturbative effects
\cite{Neubert:1993ch,Bigi:1993ex}. Recently, these non-local
structures have been analyzed systematically beyond the leading order
in $\Lambda/m_b$ \cite{Lee:2004ja,Bosch:2004cb,Beneke:2004in}. It is
generally believed that the non-local operators reduce to local ones
when the differential decay distributions are integrated over all of
phase-space. Here we show that this is not always the case.

A precise control of hadronic power corrections is particularly
important in the case of the inclusive radiative decay $\bar B\to
X_s\gamma$, which is the prototype of all flavor-changing neutral
current processes. A significant effort is currently underway to
complete the calculation of the leading-power (in $\Lambda/m_b$)
contribution to the decay rate at next-to-next-to-leading order in
renormalization-group improved perturbation theory. This leaves
non-perturbative power corrections as the potentially largest source
of theoretical uncertainty.

It is well-known that in $\bar B\to X_s\gamma$ decay the OPE faces
some limitations, which result from the fact that the photon has a
partonic substructure. For instance, there exists a contribution to
the total decay rate involving the interference of the $b\to s\gamma$
transition amplitude mediated by the electro-magnetic dipole operator
$Q_{7\gamma}$ with the charm-penguin amplitude ($b\to c\bar c s$
followed by $c\bar c\to\gamma g$) mediated by the current-current
operator $Q_1$ (see \cite{Beneke:2001ev} for the definition of the
operators in the effective weak Hamiltonian). When the charm-quark is
treated as a heavy quark ($m_c\sim m_b$), this contribution can be
expanded in local operators
\cite{Voloshin:1996gw,Ligeti:1997tc,Grant:1997ec,Buchalla:1997ky}, and
it is believed to be a good approximation to keep only the first term
in this expansion. Its contribution to the total $\bar B\to X_s\gamma$
decay rate can be written as
\[
   \frac{\Delta\Gamma}{\Gamma_{77}}
   = - \frac{C_1}{C_{7\gamma}}\,\frac{\lambda_2}{9m_c^2}
   \approx 0.03 \,,
\]
where $\lambda_2=\frac14(m_{B^*}^2-m_B^2)$, and 
\[
   \Gamma_{77} = \frac{G_F^2\alpha}{32\pi^4}\,|V_{tb} V_{ts}^*|^2 m_b^5\,
   |C_{7\gamma}|^2
\]
is the leading-order contribution to the decay rate from the
electro-magnetic dipole operator. The ratio $\Delta\Gamma/\Gamma_{77}$
therefore provides an estimate of the relative magnitude of the
non-perturbative effect. We stress that when the scaling
$m_c^2\sim\Lambda m_b$ is adopted instead of $m_c\sim m_b$, then the
charm-loop contribution must be described by the matrix element of a
non-local operator
\cite{Ligeti:1997tc,Grant:1997ec,Buchalla:1997ky,inprep}.

It has been noted in \cite{Ligeti:1997tc} that the OPE for $\bar B\to
X_s\gamma$ decay breaks down when one includes diagrams from operators
other than $Q_{7\gamma}$, in which the photon couples to light
quarks. An example of such a contribution, resulting from the decay
$b\to sg$ mediated by the chromo-magnetic dipole operator $Q_{8g}$
followed by photon emission from the light partons, was studied in
\cite{Kapustin:1995fk}. It was argued that the corresponding effect
can be estimated in terms of the parton fragmentation functions of a
quark or gluon into a photon. Since this contribution is numerically
very small, it has not received much further attention in the
literature. A more careful analysis reveals that the correct
interpretation of this effect is in terms of a subleading shape
function \cite{inprep}.

In this Letter, we identify and analyze a novel class of non-local
power corrections to the $\bar B\to X_s\gamma$ decay rate, which were
not considered before. We argue that they can affect the total decay
rate at the few percent level, and that they give the dominant
flavor-specific contribution to the rate difference between charged
and neutral $B$ mesons. The presence of this effect leads to a
dominant and irreducible source of theoretical uncertainty in the
prediction for the total $\bar B\to X_s\gamma$ branching fraction.

\section{Non-local power corrections}

\begin{figure}
\epsfxsize=8.3cm
\centerline{\epsffile{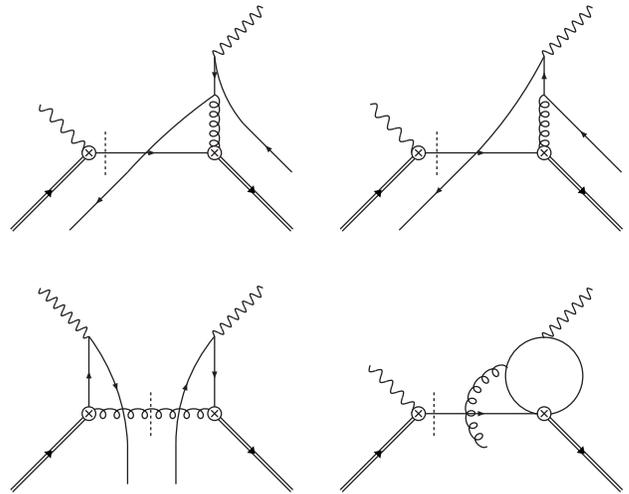}}
\vspace{0.2cm} \centerline{\parbox{14cm}{\caption{\label{fig:graphs}
Diagrams representing enhanced non-local power corrections to the
$\bar B\to X_s\gamma$ photon spectrum, which do not reduce to local
power corrections to the total decay rate. The double lines represent
heavy-quark fields $h_v$. The vertical dashed lines indicate cuts of
the relevant propagators.}}}
\end{figure}

Power corrections to the high-energy part of the $\bar B\to X_s\gamma$
photon spectrum can be systematically parameterized in terms of
subleading shape functions defined in terms of forward $B$-meson
matrix elements of non-local light-cone string operators
\cite{Lee:2004ja,Bosch:2004cb,Beneke:2004in}. Some of these operators
-- the ones considered so far in the literature -- reduce to local
operators when one considers the total decay rate (i.e., the integral
over the photon spectrum); however, a detailed analysis shows that
several of them do not \cite{inprep}. Representative diagrams giving
rise to such operators are depicted in Figure~\ref{fig:graphs}. The
graphs show different contributions to the discontinuity of the
hadronic tensor $W^{\mu\nu}$, which determines the $\bar B\to
X_s\gamma$ photon-energy spectrum via the optical theorem. The total
decay rate is obtained by an integration over the photon energy. In
this Letter we focus on the two graphs shown in the first row (and two
mirror graphs, in which the order of the weak vertices is
interchanged).  They describe the interference of the $b\to s\gamma$
transition amplitude mediated by the electro-magnetic dipole operator
$Q_{7\gamma}$ with the $b\to sg$ amplitude mediated by the
chromo-magnetic dipole operator $Q_{8g}$ followed by the fragmentation
of the gluon into an energetic photon and a soft quark-antiquark
pair. While other diagrams, such as the first graph in the second row
in the figure, give rise to four-quark operators containing strange
quarks, the graphs in the first row produce all light-quark
flavors. We expect that the resulting four-quark operators will have a
larger overlap with the $B$-meson states and thus give rise to the
dominant power corrections. For simplicity, we also do not consider
loop-suppressed effects such as the second graph in the second row of
the figure. This diagram would match onto a non-local operator
containing a soft gluon field, which mixes with the operators we
consider under renormalization.

The top two diagrams in Figure~\ref{fig:graphs} affect the $\bar B\to
X_s\gamma$ photon spectrum in the high-energy region, $E_\gamma\approx
m_b/2$, where it is most accessible to experiment. This is enforced by
the fact that the amplitude mediated by the insertion of $Q_{8g}$
interferes with the two-body decay amplitude mediated by the insertion
of $Q_{7\gamma}$. The effect can therefore not be eliminated using
kinematical cuts. We find that the corresponding contribution to the
total decay rate can be parameterized in terms of forward $B$-meson
matrix elements of tri-local light-cone operators (with $C_F=4/3$ for
$N_c=3$ colors):
\begin{eqnarray}\label{DGamma}
   \Delta\Gamma
   &=& - \Gamma_{77}\,\frac{C_{8g}}{C_{7\gamma}}\,\frac{4\pi\alpha_s}{N_c m_b}
    \int_{-\infty}^0\!ds \int_{-\infty}^0\!dt \nonumber\\
   &\times& \langle\bar B|\,C_F\,(O_1+O_2) - (T_1+T_2)\,|\bar B\rangle \,,
\end{eqnarray}
where we have used that the Wilson coefficients $C_i$ are real in the
Standard Model. The relevant factorization scale to use in this result
is of order $\mu^2\sim m_b\Lambda$. We use a mass-independent
normalization of meson states, such that $\langle\bar B|\,\bar h_v h_v
|\bar B\rangle=2$. The four-quark operators are defined as
\begin{eqnarray}
   O_1 &=& \sum_q e_q\,\bar h_v(0)\,\Gamma_R\,q(t\bar n)\,\,
    \bar q(s\bar n)\,\Gamma_R\,h_v(0) \,, \nonumber\\
   O_2 &=& \sum_q \frac{e_q}{2}\,
    \bar h_v(0)\,\Gamma_R\gamma_{\perp\alpha}\,q(t\bar n)\,\,
    \bar q(s\bar n)\,\gamma_\perp^\alpha\Gamma_R\,h_v(0) \,, \nonumber\\
   T_1 &=& \sum_q e_q\,\bar h_v(0)\,\Gamma_R\,t_a  q(t\bar n)\,\,
    \bar q(s\bar n)\,\Gamma_R\,t_a h_v(0) \,, \nonumber\\
   T_2 &=& \sum_q \frac{e_q}{2}\,
    \bar h_v(0)\,\Gamma_R\gamma_{\perp\alpha}\,t_a q(t\bar n)\,\,
    \bar q(s\bar n)\,\gamma_\perp^\alpha\Gamma_R\,t_a h_v(0) \,, \nonumber
\end{eqnarray}
where $e_q$ is the electric charge of the soft light quark in units of
$e$, $\Gamma_R=\nbslash\,(1+\gamma_5)/2$ is a right-handed Dirac
structure, and $t_a$ are the generators of color SU(3). Here $h_v$ are
the two-component heavy-quark fields defined in heavy-quark effective
theory, while $q$ are soft light-quark fields ($q=u,d,s$). The
light-quark fields are located on the light-cone defined by the
direction of the emitted photon with momentum $q^\mu=E_\gamma\bar
n^\mu$ (with $\bar n^2=0$). The non-local operators are made gauge
invariant by the insertion of soft Wilson lines $S_{\bar n}$ in the
$\bar n$-direction. The Wilson lines are absent in light-cone gauge
$\bar n\cdot A=0$, which we adopt implicitly to simplify the notation.

\begin{figure}
\epsfxsize=5.45cm
\centerline{\epsffile{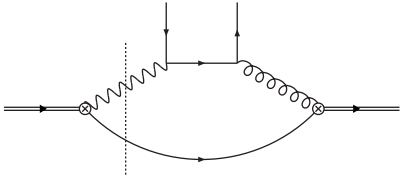}}
\vspace{0.2cm}
\centerline{\parbox{14cm}{\caption{\label{fig:cut}
Diagram representing an enhanced non-local power correction to the total 
$\bar B\to X_s\gamma$ decay rate.}}}
\end{figure}

Evaluating the hadronic matrix elements in (\ref{DGamma}) using a
systematic non-perturbative approach is a very challenging task. In
particular, lattice QCD is unable to handle operators with component
fields separated by light-like distances. Naive dimensional analysis
suggests that
$\Delta\Gamma/\Gamma_{77}\sim(C_{8g}/C_{7\gamma})\,\pi\alpha_s(\Lambda/m_b)$,
which could easily amount to a 5\% correction to the decay rate. In
more traditional applications of the OPE to inclusive $B$-meson
decays, four-quark operators contribute at order $(\Lambda/m_b)^3$ in
the heavy-quark expansion.  The non-local operators in (\ref{DGamma})
lead to enhanced power corrections of order $\Lambda/m_b$, because the
two ``vertical'' propagators in Figure~\ref{fig:graphs} have
virtualities of order $m_b\Lambda$ and so introduce two powers of soft
scales in the denominator. This mechanism was first studied in
\cite{Hill:2002vw}. The existence of such enhanced power corrections
in the total decay rate may seem puzzling at first sight. Consider the
diagram in Figure~\ref{fig:cut}, which represents the contribution to
the total rate derived from the first graph in
Figure~\ref{fig:graphs}. Before taking the cut indicated by the
vertical dashed line the diagram only receives contributions from hard
loop momenta $p^\mu\sim m_b$, and it would thus seem appropriate to
shrink all propagators to a point. In this case the diagram would
contribute at order $(\Lambda/m_b)^3$ in the heavy-quark expansion,
and this scaling would appear to be preserved when one takes the
discontinuity of the diagram, apparently contradicting our
conclusion. The loophole is that the contribution to the total $\bar
B\to X_s\gamma$ decay rate is not given by the discontinuity of the
loop graph, which would correspond to the sum of all three possible
cuts, but instead it is given by the single cut shown in the figure.

In order to obtain at least some model estimate of the magnitude of
the effect in (\ref{DGamma}) we adopt the vacuum insertion
approximation (VIA), in which the vacuum state $|0\rangle\langle 0|$
is inserted between the two light-quark fields inside the four-quark
operators. This is a crude approximation, which however appears to
work well in the analysis of $b$-hadron lifetimes
\cite{Bigi:1994wa,Neubert:1996we}. The approximation has thus been
checked for local four-quark operator evaluated between $B$-meson
states. Also, it can be justified using large-$N_c$ counting
rules. Applications of the VIA to non-local operators can be found in
\cite{Beneke:2004in,Neubert:2004cu}

In the present case, the matrix elements of the operators $O_2$ and
$T_{1,2}$ vanish in the VIA, either due to the color-octet structure
of the quark bilinears ($T_{1,2}$) or due to the fact that there is no
external perpendicular Lorentz vector available ($O_2$ and $T_2$). The
matrix element of $O_1$ can be expressed in terms of the leading
light-cone distribution amplitude of the $B$-meson in position space
\cite{Grozin:1996pq,Lange:2003ff}. We obtain
\[
   \langle \bar B|\,O_1\,|\bar B\rangle_{\rm VIA}
   = e_q\,\frac{f_B^2 m_B}{4}\,\widetilde\phi_+^B(s)\,
   [\widetilde\phi_+^B(t)]^* \,,
\]
where $e_q$ now refers to the charge of the light spectator quark in the $B$ 
meson. The integral over the position-space distribution amplitude can be 
evaluated to yield
\[
   -i \int_{-\infty}^0\!ds\,\widetilde\phi_+^B(s)
   = \int_0^\infty\!\frac{d\omega}{\omega}\,\phi_+^B(\omega)
   = \frac{1}{\lambda_B} \,,
\]
where $\phi_+^B(\omega)$ is the light-cone distribution amplitude in
momentum space, and $\lambda_B^{-1}$ is the common notation for the
first inverse moment of this quantity \cite{Beneke:2001ev}. Numerical
estimates of $\lambda_B$ are very uncertain, but typically fall in the
range between 0.25 and 0.75\,GeV
\cite{Beneke:2001ev,Grozin:1996pq,Lange:2003ff,Ball:2003fq,Braun:2003wx,Lee:2005gz}. In
the VIA, our estimate of the spectator-dependent, non-local power
corrections then takes the final form
\begin{equation}
   \frac{\Delta\Gamma_{\rm VIA}}{\Gamma_{77}}
   = - \frac{e_q C_{8g}}{C_{7\gamma}}\,
   \frac{\pi\alpha_s}{2} \left( 1 - \frac{1}{N_c^2} \right)
   \frac{f_B^2 m_B}{\lambda_B^2 m_b} \,.
\end{equation}
Recalling that $f_B\sim 1/\sqrt{m_B}$ in the heavy-quark limit, we
indeed recover the scaling behavior anticipated above. Numerically,
with $\mu\sim 1.5$\,GeV as a typical factorization scale and
$f_B\approx 0.215$\,GeV for the $B$-meson decay constant (see
\cite{Gray:2005ad} for a recent determination using unquenched lattice
QCD), we obtain
\[
   \frac{\Delta\Gamma_{\rm VIA}}{\Gamma_{77}}
   \approx - 0.26 e_q \left( \frac{f_B}{\lambda_B} \right)^2
   \approx - 0.05 e_q \left( \frac{\lambda_B}{0.5\,\mbox{GeV}} \right)^{-2} ,
\]
where $e_q=2/3$ for decays of $B^-$ mesons, while $e_q=-1/3$ for decays of 
$\bar B^0$ mesons. For the range of $\lambda_B$ values quoted above, the 
effect is between $-2\%$ and $-19\%$ times $e_q$. Taking $\Gamma_{77}$ as an 
estimate of the total decay rate at leading power, this implies that the 
enhanced power corrections to the total, flavor-averaged $\bar B\to X_s\gamma$ 
decay rate are expected (in the VIA) to be between $-0.3\%$ and $-3\%$, while 
these effects induce a flavor-dependent rate asymmetry
\begin{equation}\label{asym}
   \frac{\Gamma(B^-\to X_s\gamma)-\Gamma(\bar B^0\to X_s\gamma)}%
        {\Gamma(\bar B\to X_s\gamma)}
   \approx - 0.05 \left( \frac{\lambda_B}{0.5\,\mbox{GeV}} \right)^{-2} ,
\end{equation}
which could amount to an effect between $-2\%$ and $-19\%$.

When considering these estimates one should keep in mind that the VIA
can at best provide a very simple model of the effect of the non-local
four-quark operators in (\ref{DGamma}). Conservatively, we can
therefore not exclude that the type of enhanced power corrections
identified in this Letter could contribute to the total $\bar B\to
X_s\gamma$ decay rate at the 5\% level. The magnitude of the
flavor-specific effects studied above could be probed by a measurement
of the flavor asymmetry (\ref{asym}); but there are other four-quark
operator contributions with flavor structure $\bar b s\,\bar s b$ (see
e.g.\ the bottom left diagram in Figure~\ref{fig:graphs}), whose
matrix elements vanish in the VIA but could still be significant in
real QCD. Their contributions are flavor-blind and hence not tested by
(\ref{asym}).

The Babar collaboration has measured the flavor-dependent rate
asymmetry in eq.~(\ref{asym}), finding the value $(1.2\pm 11.6\pm
1.8\pm 4.8)\%$ , where the errors are statistical, systematic and due
to the production ratio $\bar B^0/B^-$, respectively
\cite{Aubert:2005cu}. The dominant error is statistical and therefore
likely to decrease when more data is collected.

\section{Conclusions}

We have identified a new class of enhanced power corrections to the
total inclusive $\bar B\to X_s\gamma$ decay rate, which cannot be
parameterized in terms of matrix elements of local operators. These
effects are nevertheless ``calculable'' in the sense that they can be
expressed in terms of subleading shape functions. At tree level, the
corresponding operators are tri-local four-quark operators. While
local four-quark operators contribute at order $(\Lambda/m_b)^3$ in
the heavy-quark expansion of the total decay rate, the effects we have
explored are enhanced by the non-local structure of the operators and
promoted to the level of $\Lambda/m_b$ corrections. We have identified
and estimated what we believe are the dominant corrections of this
type, namely those that match the flavor quantum numbers of the
external $B$-meson states.

Our results imply that a local operator product expansion for the
inclusive $\bar B\to X_s\gamma$ decay rate does not exist. Even at
first order in $\Lambda/m_b$ there are hadronic effects that can only
be accounted for in terms of non-local operators. The precise impact
of these power corrections will be notoriously difficult to estimate
using our present command of non-perturbative QCD on the light
cone. While a naive estimate using the vacuum insertion approximation
suggests that the effects are at the few percent level, we conclude
that they are nevertheless a source of significant hadronic
uncertainty in the calculation of partial or total $\bar B\to
X_s\gamma$ decay rates. After the perturbative analysis of the decay
rate will have been completed, the enhanced non-local power
corrections will remain as the dominant source of theoretical
uncertainty. A measurement of the flavor-dependent asymmetry
(\ref{asym}) could help to corroborate our numerical estimates of such
corrections.

\vspace{0.3cm}  
{\em Acknowledgments:\/} 
We are grateful to Thomas Becher and Martin Beneke for useful
comments. The research of S.J.L.\ and M.N.\ is supported by the
National Science Foundation under grant PHY-0355005. The research of
G.P. is supported by the Department of Energy under grant
DE-FG02-90ER40542.

\end{document}